\documentclass[twoside,slac_one]{revtex4}
\usepackage{graphicx}
\usepackage{fancyhdr}
\usepackage{amsmath} 
\usepackage{bm}
\usepackage{amsxtra}
\usepackage{amssymb}
\usepackage{amsthm}
\usepackage{latexsym}
\usepackage{lscape}

\def\to     {\!\rightarrow\!} 
\def\ee     {e^+e^-} 
\def\KK     {K^+K^-} 
\def\pipi   {\pi^+\pi^-} 
\def\pzpz   {\pi^0\pi^0} 
\def\qbar   {\overline{q}}
\def\qqbar  {q\qbar}

\def\eepipi  {$\ee \to \pipi$}
\def\eekkppc {$\ee \to \KK\pipi$}
\def\eekkppn {$\ee \to \KK\pzpz$}
\def\sqrts   {$\sqrt{s}$}
\def\sqrtsp  {$\sqrt{s^\prime}$}
\def\pz      {$\pi^0$} 
\def\etaa    {$\eta$}
\def\etap    {$\eta^\prime$}
\def\etac    {$\eta_c$}

\begin{document}

\title{\boldmath
ISR Hadron Production in $\ee$ Annihilations and 
Meson-Photon Transition Form Factors}

%

\author{D. R. Muller}
\affiliation{SLAC National Accelerator Laboratory, Stanford, CA, USA}
\affiliation{Representing the BaBar collaboration}

\begin{abstract}
We present several recent results from the BaBar collaboration in the
areas of initial state radiation physics and transition form factors.
An updated study of the processes \eekkppc\ and \eekkppn\ provides an
improved understanding of the $Y(2175)$ meson.
A very precise study of the process \eepipi\ improves the precision on
the calculated anomalous magnetic moment of the muon and provides by 
far the best information on excited $\rho$ states.
Our previous measurements of the timelike transition form factors
(TFF) of the \etaa\ and \etap\ mesons at $Q^2 \!=\! 112$~GeV$^2$,
combined with new measurements of the their spacelike TFFs and those
of the \pz\ and \etac\ mesons,
provide powerful tests of QCD and models of the distribution
amplitudes of quarks inside these mesons.
The \etac\ TFF shows the expected behavior over the $Q^2$ range
1--50~GeV$^2$, 
and we are sensitive to next-to-leading-order QCD corrections.
The \etaa\ and \etap\ TFFs are consistent with expected
behavior, but those for the \pz\ are not.
Extracting the strange and nonstrange components of the
\etaa\ and \etap\ TFFs, we find the nonstrange component to be
consistent with theoretical expectations and inconsistent with the
measured \pz\ TFF.
\end{abstract}

\maketitle

\thispagestyle{fancy}


\section{Introduction}
The BaBar experiment studies electron-positron annihilations at a
center-of-mass (CM) energy \sqrts$=$10.6~GeV.
It features a boosted CM system and a state of the art, asymmetric
detector, which were designed for the study of CP violation in the $B$
meson system at the very high luminosity PEP-II $B$ factory.
These features also enable a wide range of additional physics at and
below the nominal \sqrts.
Here we consider recent results in two such areas:
initial state radiation (ISR), which gives access to $\ee$
annihilations at lower CM energies;
and two-photon collisions, which can produce final states with quantum
numbers such as $J^{PC}=0^{-+}$ that are inaccessible via annihilations.

\section{Initial State Radiation}
An incoming electron or positron can radiate an energetic photon before
annihilating with an oncoming positron or electron, respectively, 
at a reduced CM energy \sqrtsp$=$\sqrts$-E_\gamma$
where $E_\gamma$ is the energy of the radiated photon.
This process is well understood theoretically, 
so that cross sections for and characteristics of particular types of
events can be measured over a wide range of \sqrtsp\ in a
single experiment.
If the ISR photon is well within the detector acceptance, 
as it is about 12\% of the time in BaBar, 
then the hadronic system is also well contained, 
with full acceptance in all kinematic variables.
Furthermore, the hadronic system is boosted, allowing good
measurements of kinematic quantities for energies all the way down to
threshold.
However, the ISR cross section is low, 
so that high luminosity is required to make meaningful measurements.

In BaBar, we have studied a large number of exclusive hadronic final states
produced via ISR.
In each case, we select events with a high-energy photon candidate recoiling
against a particular number of charged tracks and additional photons,
identify the tracks as pions, kaons or protons, 
and combine photons to form \pz\ candidates.
We then subject the set of reconstructed particles to a number of kinematic
fits that impose 4-momentum conservation under various hypotheses for
the event type.
We select events with a good $\chi^2$ for the signal hypothesis 
and poor $\chi^2$ for certain alternative hypotheses,
evaluate remaining backgrounds from the data and subtract them.
We derive cross sections as functions of \sqrtsp\ using efficiencies
derived from the data,
and study the structure of the events, in particular any resonant
contributions.
Here we present updated results on the $\KK\pipi$ and $\KK\pzpz$ final
states,
and new results on the $\pipi$ final state.

\subsection{\boldmath The \eekkppc\ and \eekkppn\ Processes}
We have updated our ealier study~\cite{isrkkppold} of the final states
comprising two charged kaons and either two charged or two neutral
pions using a larger data sample~\cite{isrkkppnew}.
The largest backgrounds are from other ISR processes with similar
kinematics, 
namely $K^0_SK^\pm\pi^\mp$, $\pipi\pipi$, $\pipi\pzpz$, $\KK\pi^0$, $\KK\eta$
and $\KK\pzpz\pi^0$,
all of which we have measured previously.
The backgrounds from $\ee\to\qqbar\to\pi^0\KK\pi\pi$, 
where an isolated, energetic $\pi^0$ mimics an ISR photon,
are also substantial.
We measure them in the data by combining the ISR
photon with additional energy clusters in the event and fitting the
\pz\ peaks in the resulting invariant mass distributions.
These backgrounds are at the few percent level at low \sqrtsp,
but become relatively large at higher energies, 
limiting the range of our measurements to about 5~GeV.

\begin{figure*}[ht]
\centering
\includegraphics[width=84mm]{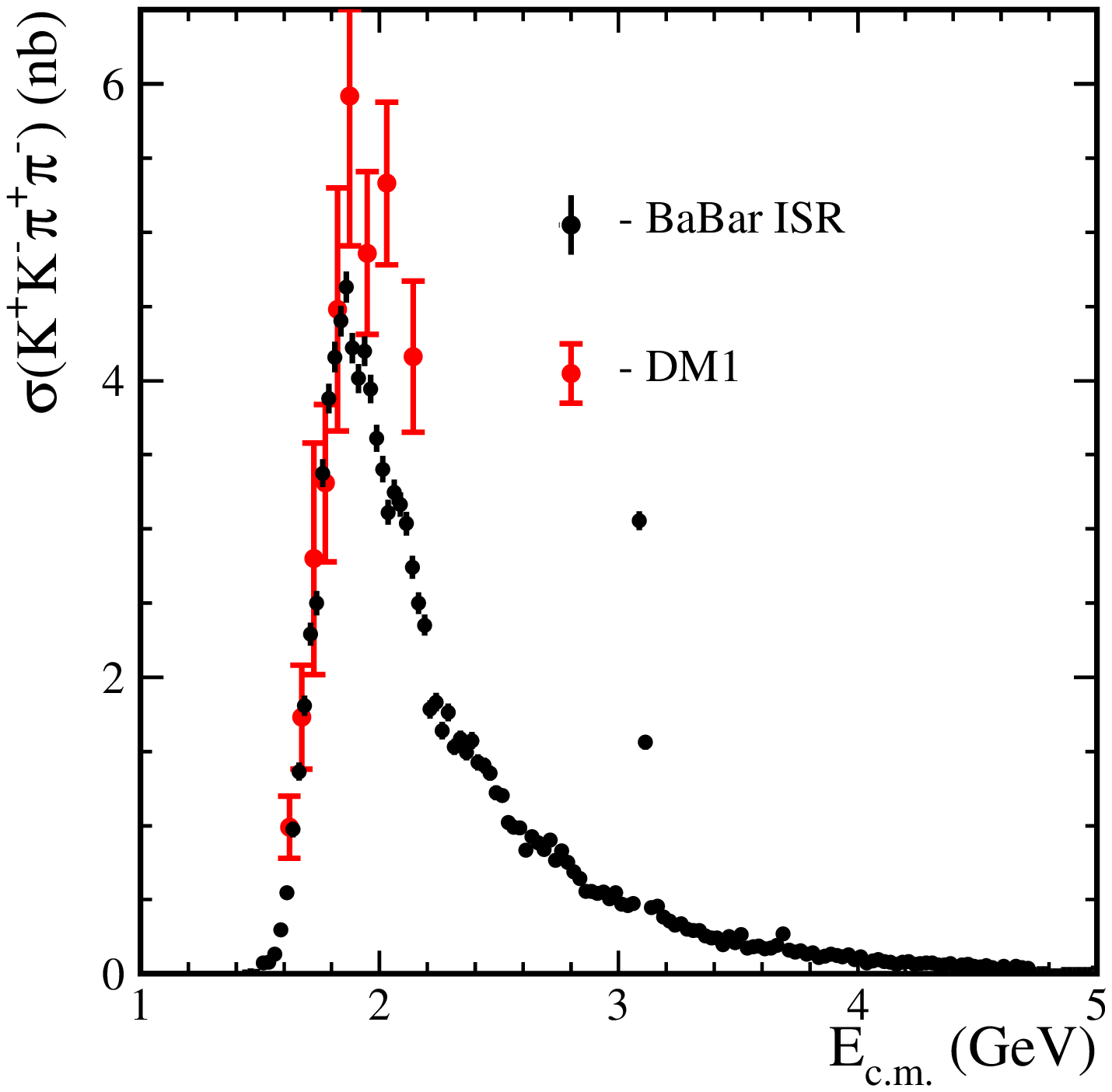}
\includegraphics[width=84mm]{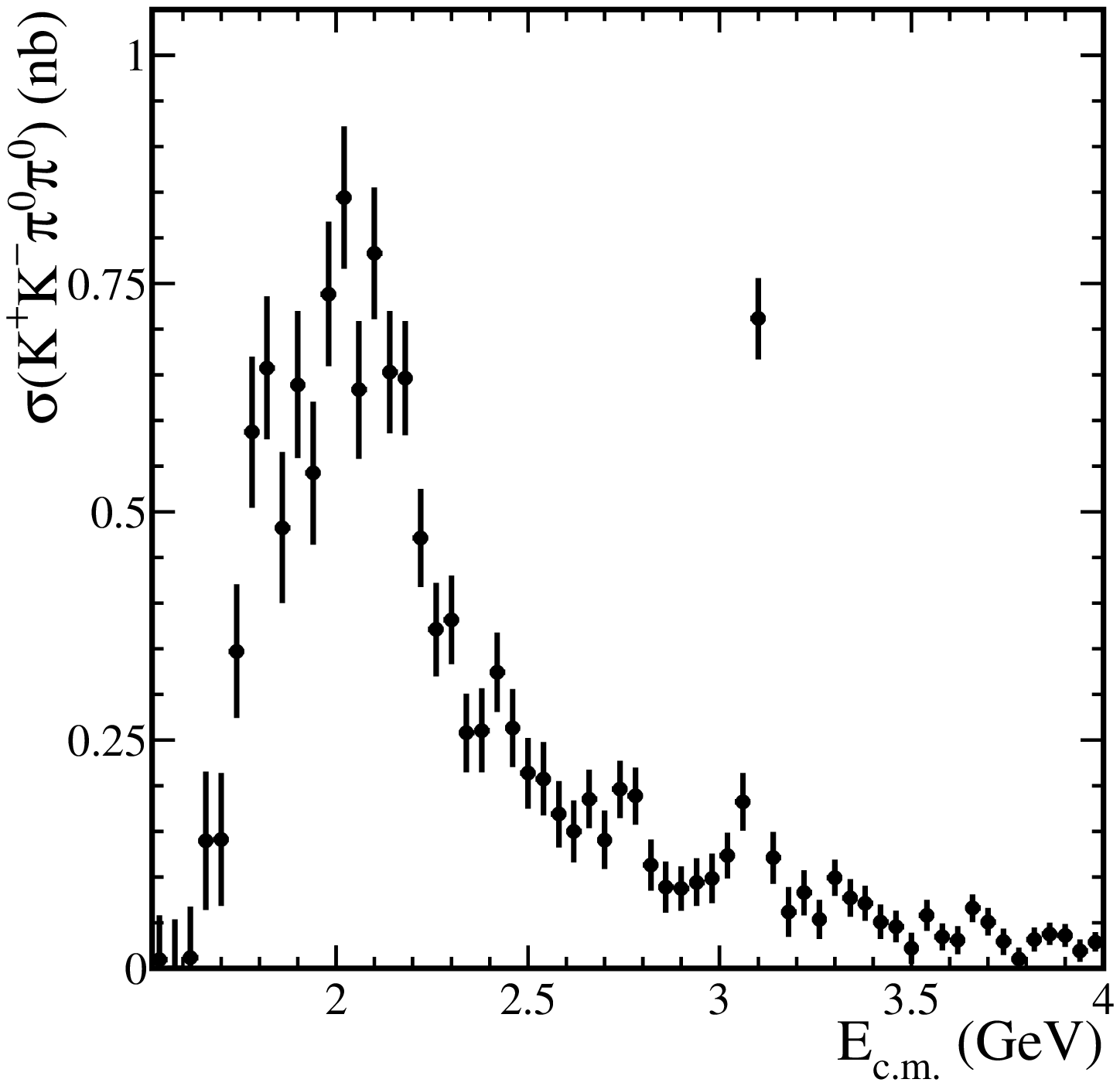}
\caption{Cross sections for the processes \eekkppc (left) and \eekkppn
  (right) as functions of the $e^+e^-$ CM energy $E_{c.m.}\!=$\sqrtsp.
  Previous results from the DM1 experiment are also shown, in red.}
\label{isrkkppxsecs}
\end{figure*}

We calibrate the efficiencies for particle reconstruction and
identification from the data, 
and use them to convert the observed numbers of events in each \sqrtsp\ bin
into the cross sections shown in Fig.~\ref{isrkkppxsecs}.
These supersede our previous measurements.
The $\KK\pipi$ measurement is consistent with and far more precise
than the only other measurement, from the DM1 collaboration~\cite{dm1}, 
for energies below 2.2~GeV.
Ours remains the only measurement at higher energies, 
extending to 5~GeV, 
and our measurement of the $\KK\pzpz$ cross section from threshold to
4~GeV is also still unique.
The errors shown are statistical;  
there is an overall 5\% (7\%) systematic uncertainty on the $\KK\pipi$
($\KK\pzpz$) cross section.

Both cross sections are very small from threshold up to about
1.6~GeV,
at which point they rise quickly to peak values of about 5~nb and 0.8~nb,
and then fall with increasing \sqrtsp.
A narrow $J/\psi$ peak is clear in both final states, 
and a $\psi(2S)$ peak is visible in the $\KK\pipi$ mode.
The $\KK\pipi$ cross section also shows considerable structure in the
1.8--3~GeV region, 
and similar features may be present in the $\KK\pzpz$ cross section,
though they cannot be resolved with the current statistics.

\begin{figure*}[ht]
\centering
\includegraphics[width=56mm]{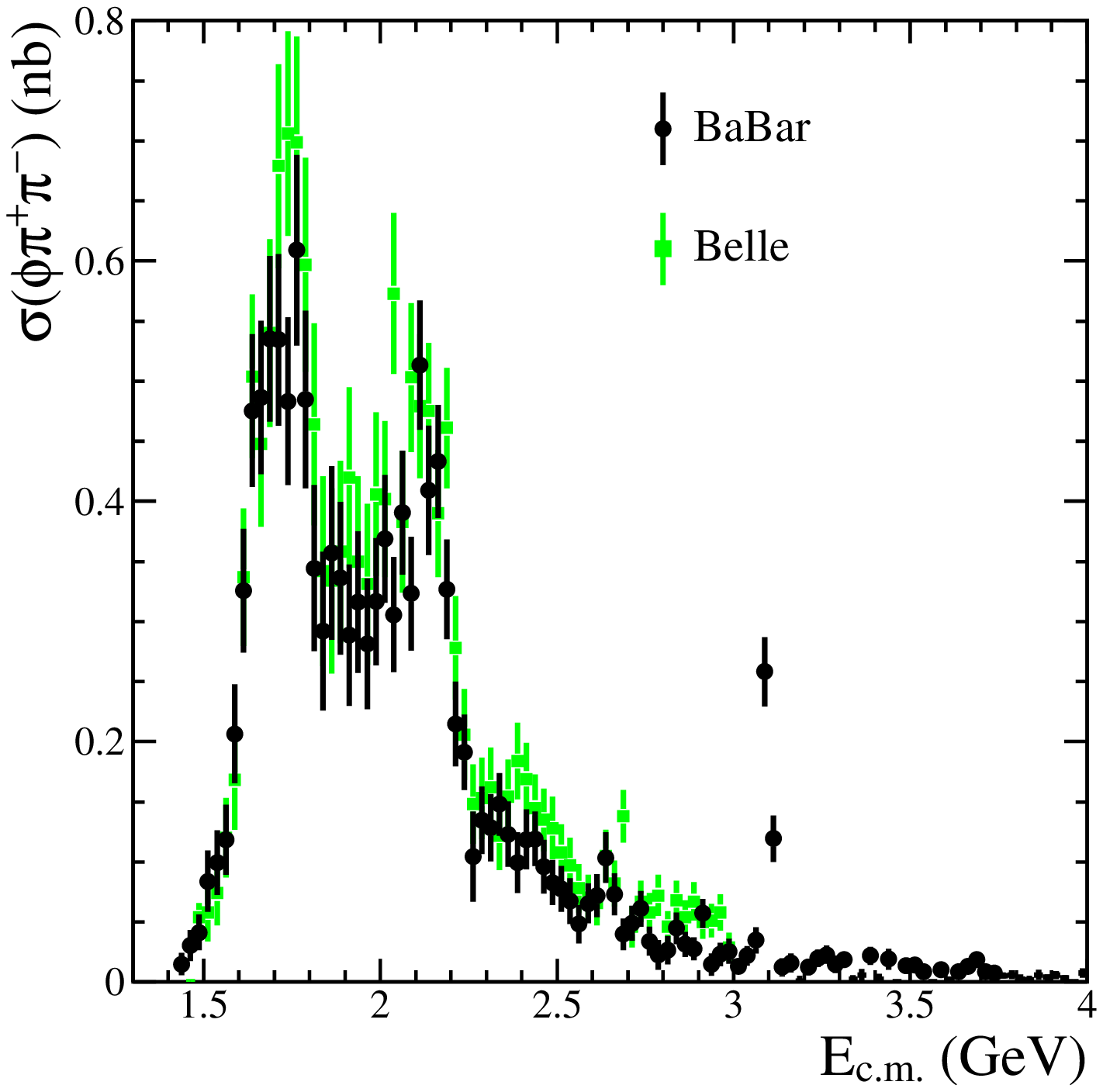}
\includegraphics[width=56mm]{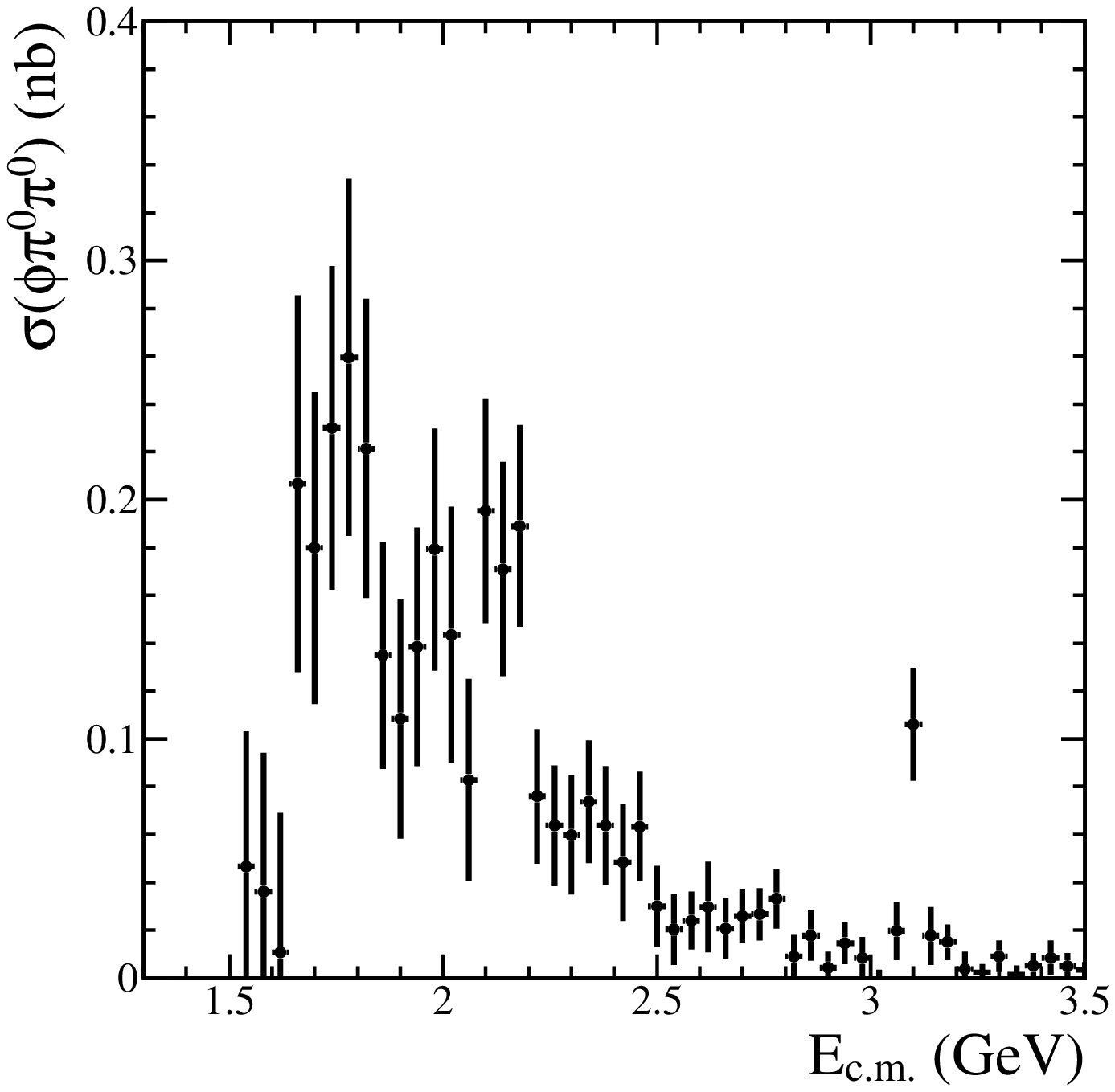}
\includegraphics[width=56mm]{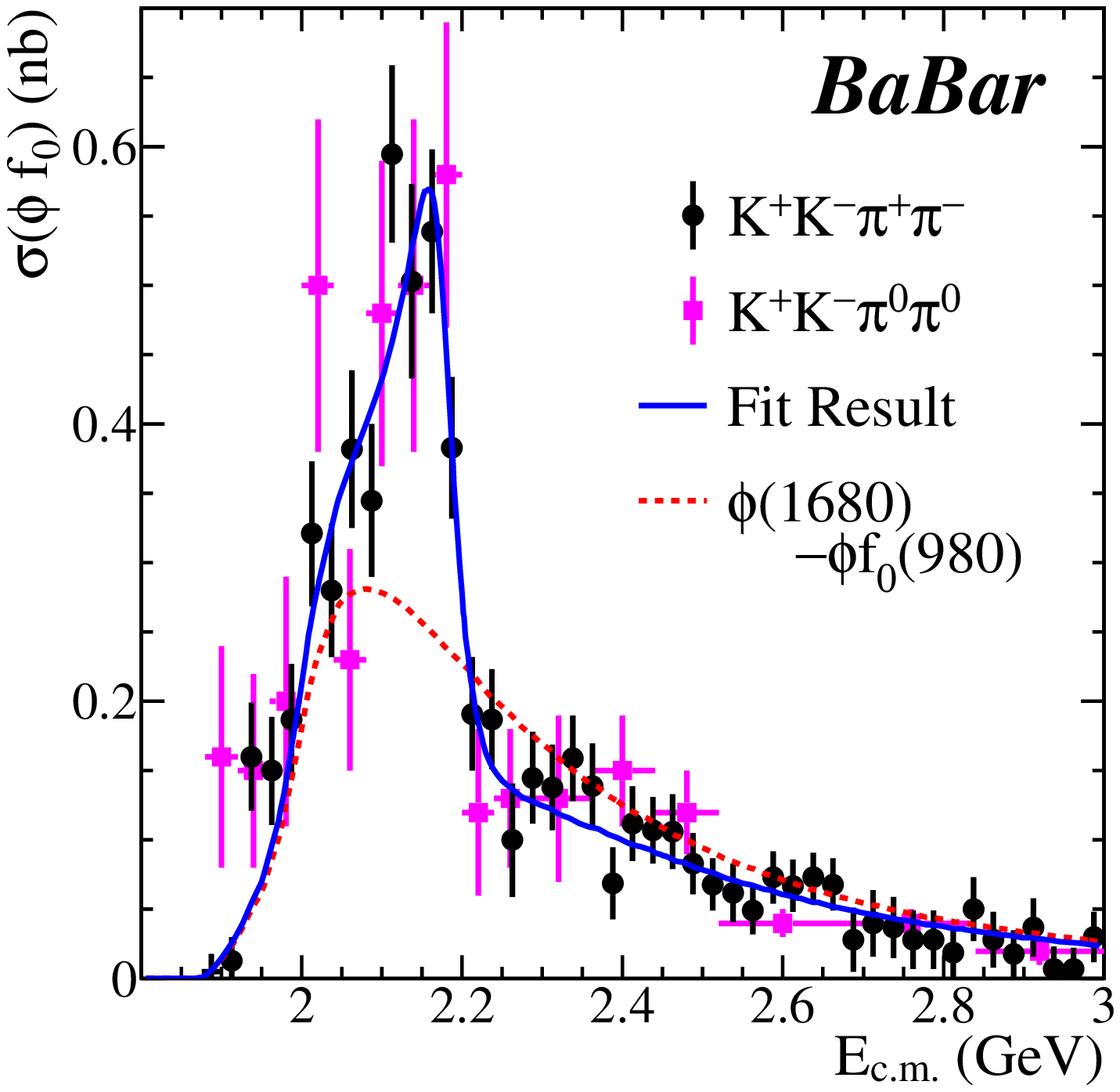}
\caption{
Cross sections for the processes $\ee\to\phi\pipi$ (left) and
$\ee\to\phi\pzpz$ (middle) vs.\ \sqrtsp.
The right plot shows the cross sections for $\ee\to\phi f_0$
measured in the $\KK\pipi$ (black) and $\KK\pzpz$ (magenta) final
states.
The blue line is the result of the fit described in the text,
with the dashed red line indicating the $\phi(1680)$ component.} 
\label{isrphippxsecs}
\end{figure*}

Both final states are dominated by the quasi-3-body submodes
$K^*(890)K\pi$, $K^*(1430)K\pi$, $\KK\rho$ and $\phi\pi\pi$.
There are substantial contributions from the quasi-2-body modes
$K^*(890)\overline{K}^*(890)$, $K^*(890)\overline{K}^*(1430)$, 
$K_1^+(1270)K^-$ and $\phi f_0(980)$.
We derive cross sections for most of these modes individually by
fitting sets of invariant mass distributions.
The $\phi\pi\pi$ and $\phi f_0(980)$ contributions are particularly
interesting, and their cross sections are shown in 
Fig.~\ref{isrphippxsecs}.
The peaks near threshold in the $\phi\pi\pi$ cross sections (left and
middle plots) are expected from the $\phi(1680)$ resonance.
The peaks at higher masses are from the $Y(2175)$, 
a state first reported by us~\cite{isrkkppold}, 
since confirmed by the BES~\cite{bes} and 
Belle (as shown in Fig.~\ref{isrphippxsecs})~\cite{belle}
experiments, 
and seen in this study with a significance exceeding nine standard
deviations.

These cross sections can be described using only two contributions, 
a $\phi(1680)$ that decays into both $\phi\pi\pi$ and $\phi f_0$, 
and a $Y(2175)$ that decays only into $\phi f_0$.
A combined fit to these and other cross sections yields improved
measurements of the mass and width of the $Y(2175)$, 
as well as its production cross section and phase with respect to the
$\phi(1680)$:
\begin{equation}
\begin{array}{ccr@{\pm}r@{\pm}rl}
m_Y        & = &  2180 &    8 &    8 & {\rm MeV/c}^2;  \\
\Gamma_Y   & = &    77 &   15 &   15 & {\rm MeV;}      \\
\sigma_Y   & = &    93 &   21 &   10 & {\rm pb;}       \\
\psi_Y     & = & -2.11 & 0.24 & 0.12 & {\rm rad.}    \nonumber 
\end{array}\label{yparams}
\end{equation}

The result of the fit is shown as the blue line on the righ-hand plot
in Fig.~\ref{isrphippxsecs}.
Here, the $\phi f_0$ cross sections measured in the two final states
are overlaid and seen to be consistent.
The contribution from the $\phi (1680)$, shown in red,
is well constrained by the $\phi\pi\pi$ cross sections,
and the sharp drop near 2.2~GeV is well described by destructive
interference between the two resonances.

The interpretation of this state remains unclear.
It is unlikely to be an excited $\phi$ state, 
since we observe no signal when the $\pipi$ pair is outside the $f_0$
region.
It could be a strangeonium-like state analogous to one the many
recently discovered charmonium-like states.

\subsection{\boldmath The \eepipi\ Process}
Recently, 
we have completed a precise measurement of the final state comprising
only two charged pions~\cite{isrpipi}.
Here we aim for precision better than 1\%, 
so the analysis is more detailed.
The largest backgrounds are from ISR $\mu^+\mu^-$ and ISR $\KK$ production, 
and we measure these two processes simultaneaously with the signal.
Since there are \sqrtsp\ regions in which each of these dominates, 
we are able to calibrate all particle identification and
misidentification rates reliably from the data.
We measure the backgrounds from $\ee \to \qqbar \to \pi^0 \pipi$ and
$\pi^0\KK$ from the data, as described above,
and take those from other ISR processes from our previous measurements.
The latter backgrounds are very low below about 1.4~GeV,
but become important rapidly at higher \sqrtsp.

We measure track finding efficiencies separately for muons, pions, and
kaons, 
as well as the correlated efficiency for the two tracks in a pair, 
which is driven by their proximity halfway through the tracking
volume.
Similarly, we measure track identification efficiencies and their
correlations in the data, 
with the correlations driven by dead and inefficient regions of the
muon and hadron identification systems.
We also measure the trigger efficiencies using redundant triggers.

The effects of higher order initial and final state radiation (FSR) 
are important.
We study these by considering each additional energy cluster in each
selected event as an FSR candidate
and performing a kinematic fit under this hypothesis.
We also consider the hypothesis of an additional ISR photon emitted
along the beam line and not detected.
From the distributions of the changes in $\chi^2$ for these various
additionas,
we measure the product of higher order effects and our acceptance for
such events, 
which we find to be consistent with our simulation.
This study also gives additional constraints on several of the important
backgrounds.

\begin{figure}[ht]
\centering
\includegraphics[width=170mm]{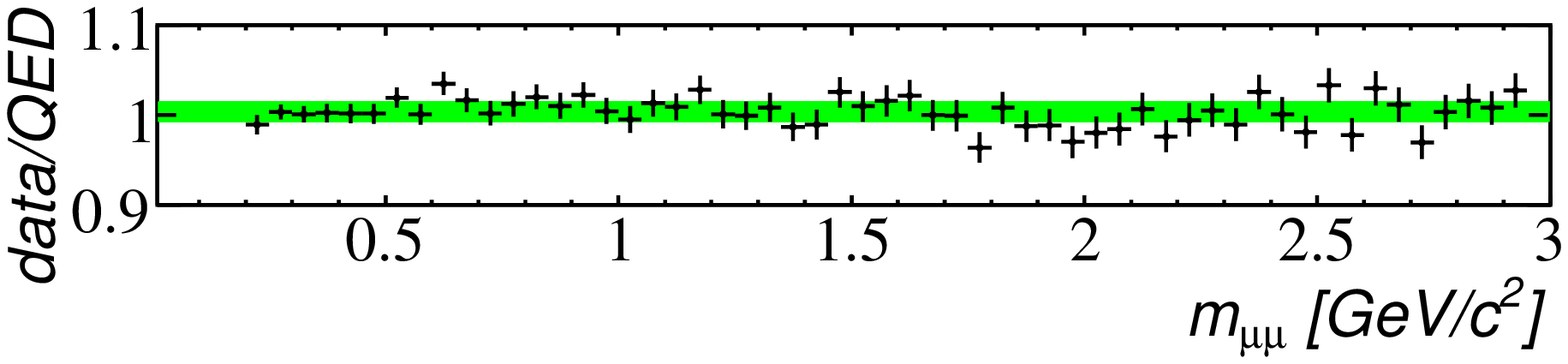}
\includegraphics[width=170mm]{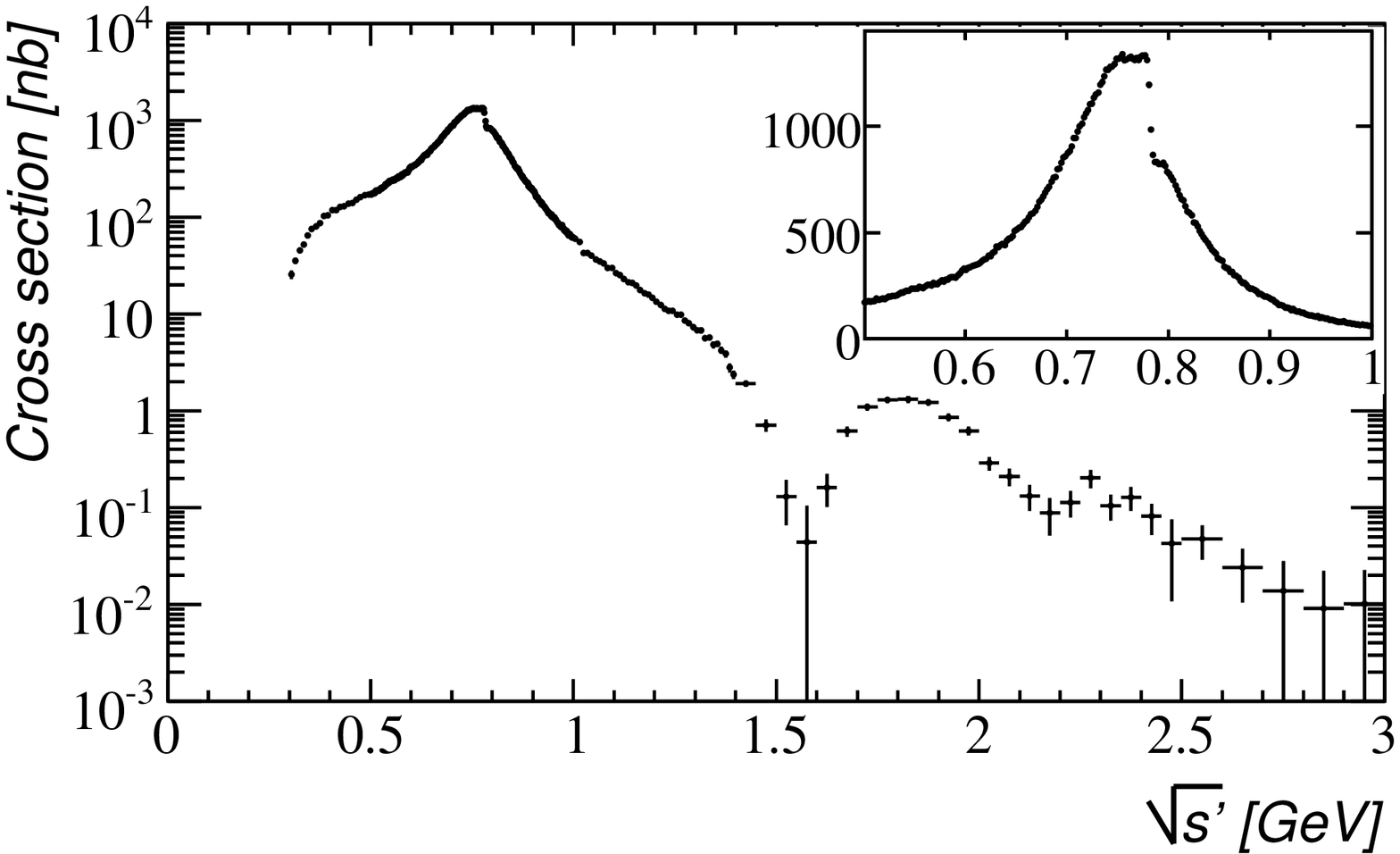}
\caption{Top:  ratio of the measured $\mu$-pair cross section to the
  prediction of QED.  The data points are shown with statistical
  errors and the green band represents the systematic uncertainty.
  Bottom:  the cross section for the process \eepipi\ as a function of
  $e^+e^-$ CM energy.} 
\label{isrpipixsec}
\end{figure}

After verifying that our measured $\mu$-pair cross section agrees with
the predictions of QED within our overall uncertainty of better than
1.7\% (see Fig.~\ref{isrpipixsec}),
we derive our $\pipi$ cross section from the ratio of our
$\pipi$ and $\mu^+\mu^-$ measurements and QED.
This cancels or reduces several of the systematic uncertainties.
The resulting cross section is shown in Fig.~\ref{isrpipixsec} and
covers by far the widest range of any single experiment,
from threshold to 3~GeV.
In the regions below 0.6~GeV and between 1 and 1.4~GeV, 
our systematic uncertainty is about 0.8\% and we are consistent and
competitive with previous results.
We have the world-best results in the regions 0.3--0.4 and
1.4--2.2~GeV,
and the only measurement above 2.2~GeV.
We observe considerable structure above 1.2~GeV that requires at least
three excited $\rho$ states for a good description.
This measurement will increase the understanding such states dramatically.

The cross section shows a very large peak due to the $\rho(770)$, 
and its interference with the $\omega$ is clearly visible, 
with deatil shown in the inset.
In this region, our systematic uncertainty is 0.5\%,
slightly better than the 0.8\% on the next best results, 
from the CMD-2~\cite{cmd2} and KLOE~\cite{kloe} experiments.
There is also a 1.5\% measurement from the SND expermient~\cite{snd}.
Results from all experiments are consistent,
but they do show variations in both the overall normalizations and
slopes at the level of the systematic uncertainties.
As with many high precision measurements, 
the interpretation can be rather sensistive to small changes in
such variables,
and it is very important to have multiple independent measurements
with similar precision, 

In particular, 
there is a long standing discrepancy between the experimentally
measured value of the anomalous magnetic moment of the muon, 
$a_\mu \!=\! g_\mu \!-\! 2$ and its theoretical calculation.
The calculation of the hadronic loop contribution $a_\mu^{\rm had}$ 
requires the total $e^+e^- \to\;$hadrons cross section as input to a
convolution integral.
The kernel is proportional to $1/s$,
so that the low-energy cross section dominates the integral, 
and that is exclusively (predominantly) from the $\pipi$ final state below
0.6 (1.0)~GeV.

We have also measured most of the final states that contribute in the
1--3~GeV region.
Taken together, 
our measurements have improved the uncertainty on $a_\mu^{\rm had}$ by
about 20\%, 
with about half the improvement from this measurement of the $\pipi$
final state.
A current global fit~\cite{davier} gives a value of $a_\mu$ that
differs from the experimental value by 29$\pm$8$\times$10$^{-10}$, 
or 3.6 standard deviations.
These measurements will have a similar effect on the calculation of
the running fine structure constant $\alpha(M_Z)$.

\section{Transition Form Factors}
A transition form factor (TFF) $F_X(q_1^2,q_2^2)$ characterizes the
coupling of a meson $X$ to a pair of (virtual) photons with squared invariant
masses $q_1^2$ and $q_2^2$.
Its dependence on the $q_i^2$ can be related to the distribution
amplitudes of quarks within the meson $X$,
providing a test of models for such amplitudes.
Its asymptotic value when one photon is real and the other highly
virtual can be related to meson's decay constant, $f_X$,
providing a test of QCD.

\subsection{Timelike TFFs}
We have measured the cross sections for the processes
$\ee\to\gamma^*\to\eta\gamma$ and
$\ee\to\gamma^*\to\eta^\prime\gamma$~\cite{tffetat}.
In each case,
there is a real photon in the final state, i.e.\ $q_2^2=0$, 
and a virtual photon with $Q^2 \!=\! q_1^2 \!=\! 10.6^2$~GeV$^2$ 
in the propagator, 
so that these cross sections can be related to the TFFs $F_\eta(Q^2)$
and $F_{\eta^\prime}(Q^2)$.
The TFF is denoted timelike since $Q^2$ is positive, and the second
argument is dropped when $q_2^2=0$.

As in our ISR analyses, we select events with an energetic photon
recoiling against either a $\pipi\pi^0$ system or a $\pipi\eta$ system
in which the \etaa\ candidate is formed from a $\gamma\gamma$ pair or
a $\pipi\pi^0$ combination.
A set of kinematic fits combined with selection criteria optimized for
this process yields very clean samples of $\ee\to\pipi\pi^0\gamma$ and
$\ee\to\pipi\eta\gamma$ events.
Since pseudoscalar mesons cannot be produced exclusively via $e^+e^-$
annihilation, 
a peak in the $\pipi\pi^0$ ($\pipi\eta$) invariant mass spectrum at the
\etaa\ (\etap) mass must be due to non-ISR processes,
and this one is expected to dominate.
From about 20 and 45 observed events, 
we derive cross sections for $\ee\to\eta\gamma$ and 
$\ee\to\eta^\prime\gamma$ of 4.5$\pm$1.2$\pm$0.3 and
5.4$\pm$0.8$\pm$0.3~fb, respectively, 
where the first errors are statistical and the second systematic.

\subsection{Spacelike TFFs}
We have measured the cross sections for the `two-photon' processes
$\ee\to\ee\gamma^*\gamma^*\to\ee X$,
for the pseudoscalar mesons $X \!=\! \pi^0$~\cite{tffpizs}, \etaa\ and
\etap~\cite{tffetas} and \etac~\cite{tffetacs}.
Here, both the electron and positron emit a virtual photon,
the two photons interact to produce the meson $X$,
and the $e^+$ and $e^-$ remain in the final state.
We consider the `single tag' case, 
in which one photon is nearly real ($q_1^2 \approx 0$)
and the $e^+$ or $e^-$ that emitted it travels along the beam line,
and the other photon has $Q^2 \!=\! -q_2^2 > 3$~GeV so that the $e^-$ or
$e^+$ is within the detector acceptance.
These cross sections can be related to the TFFs $F_X(Q^2)$.

The event selection is similar to that for ISR or timelike TFF
measurements, 
but with an energetic $e^\pm$ in place of the high-energy photon.
Since the $e^\pm$ is charged,
we trigger on purely neutral recoil systems efficienctly and
reconstruct them well.
We select events with a well identified $e^\pm$ recoiling against a
$\gamma\gamma$ pair or a $\pipi\pi^0$, $\pipi\eta$ or $K^0_SK^+\pi^-$ system,
all well containted within the detector acceptance.
To obtain clean samples of two-photon events with well measured $Q^2$,
we make requirements on the 4-momenta of the detected $e^\pm X$
systems, 
the missing 4-momenta,
and the direction (forward or backward) of the detected $e^\pm$.
We fit the invariant mass distributions of the recoiling systems to
obtain the numbers of $X$-mesons in each bin of $Q^2$.
We cover the $Q^2$ range from 4~GeV$^2$, 
below which the efficiency is low and changing rapidly, 
to 40 or 50~GeV$^2$, above which no significant signal is seen.

These fits eliminate non-$X$ backgrounds.
We estimate the backgrounds from $\ee\to\qqbar$ events using events with
the $e^\pm$ travelling in the wrong direction, 
and subtract them;
they are very small.
We estimate the backgrounds from other two-photon processes by
reconstructing additional \pz\ candidates in the events,
and by studying the shapes of the kinematic distributions used in the 
selection for data and simulated signal-like and background-like events.
These backgrounds are also very small, but must be subtracted carefully.
We observe a total of about 14,000 $\ee\to\ee\pi^0$ events, 2,800
\etaa\ events, 5,000 \etap\ events and 530 \etac\ events.

\subsection{Results}
\begin{figure*}[ht]
\centering
\includegraphics[width=56mm]{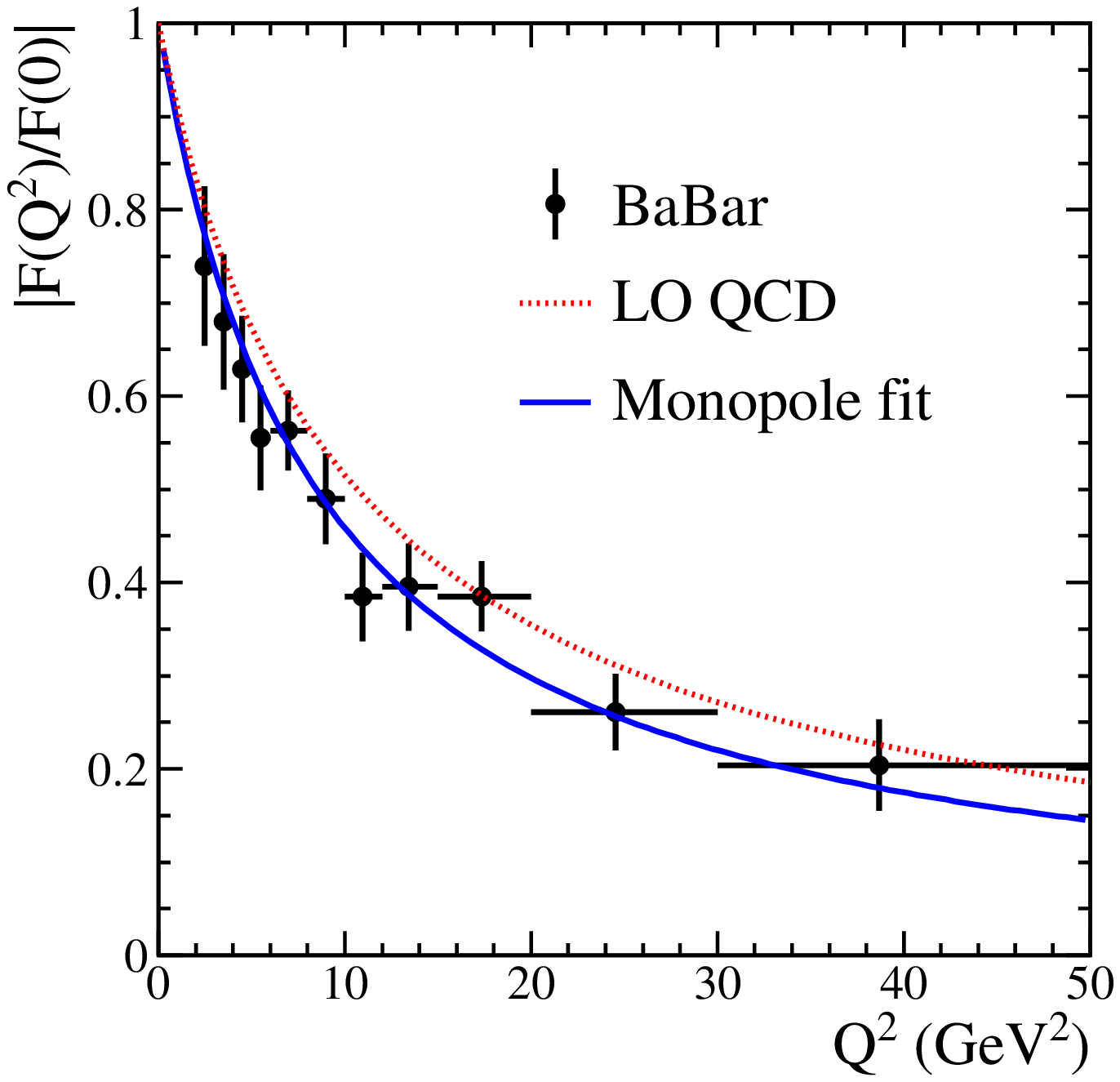}
\includegraphics[width=55mm]{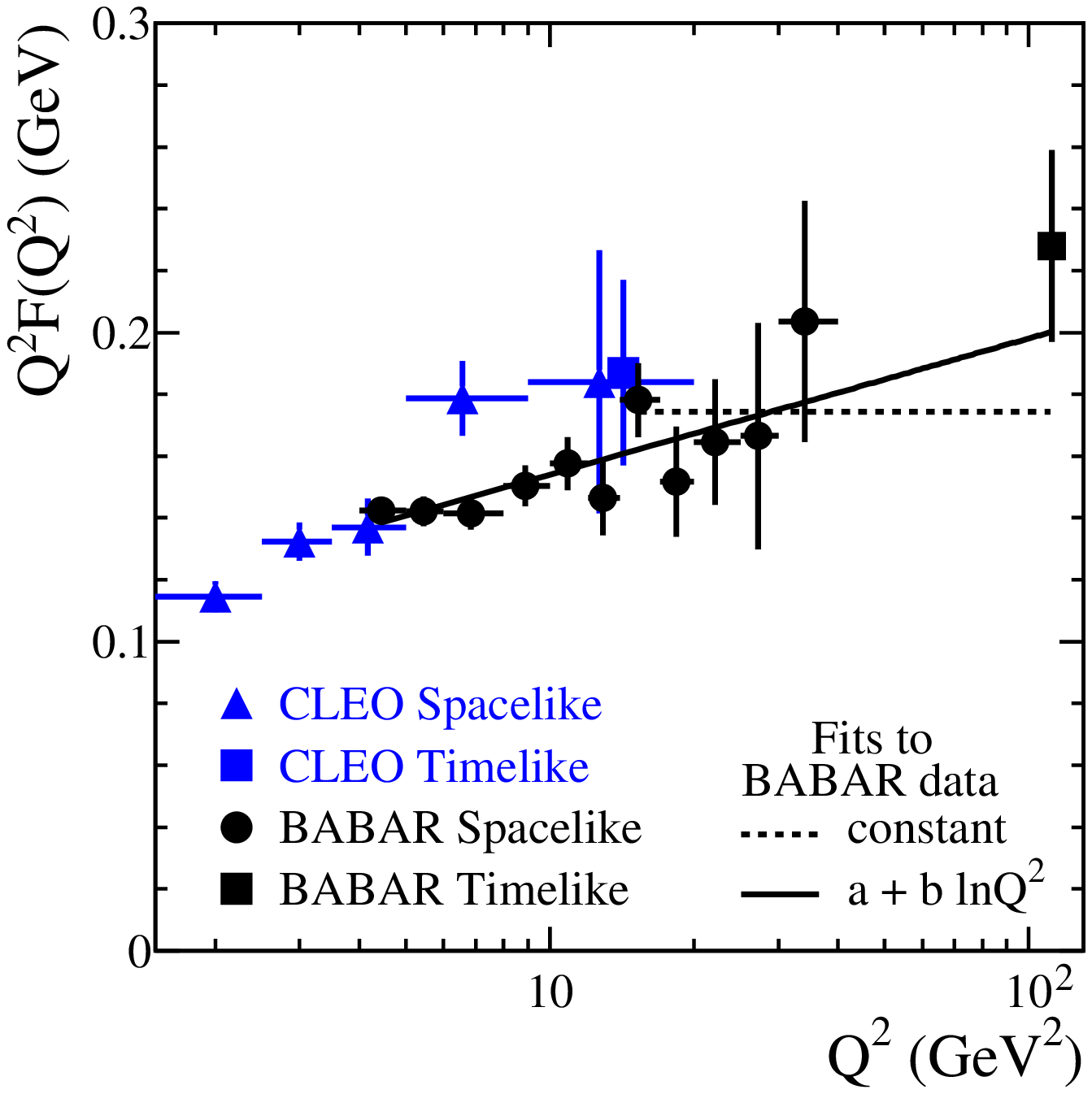}
\includegraphics[width=55mm]{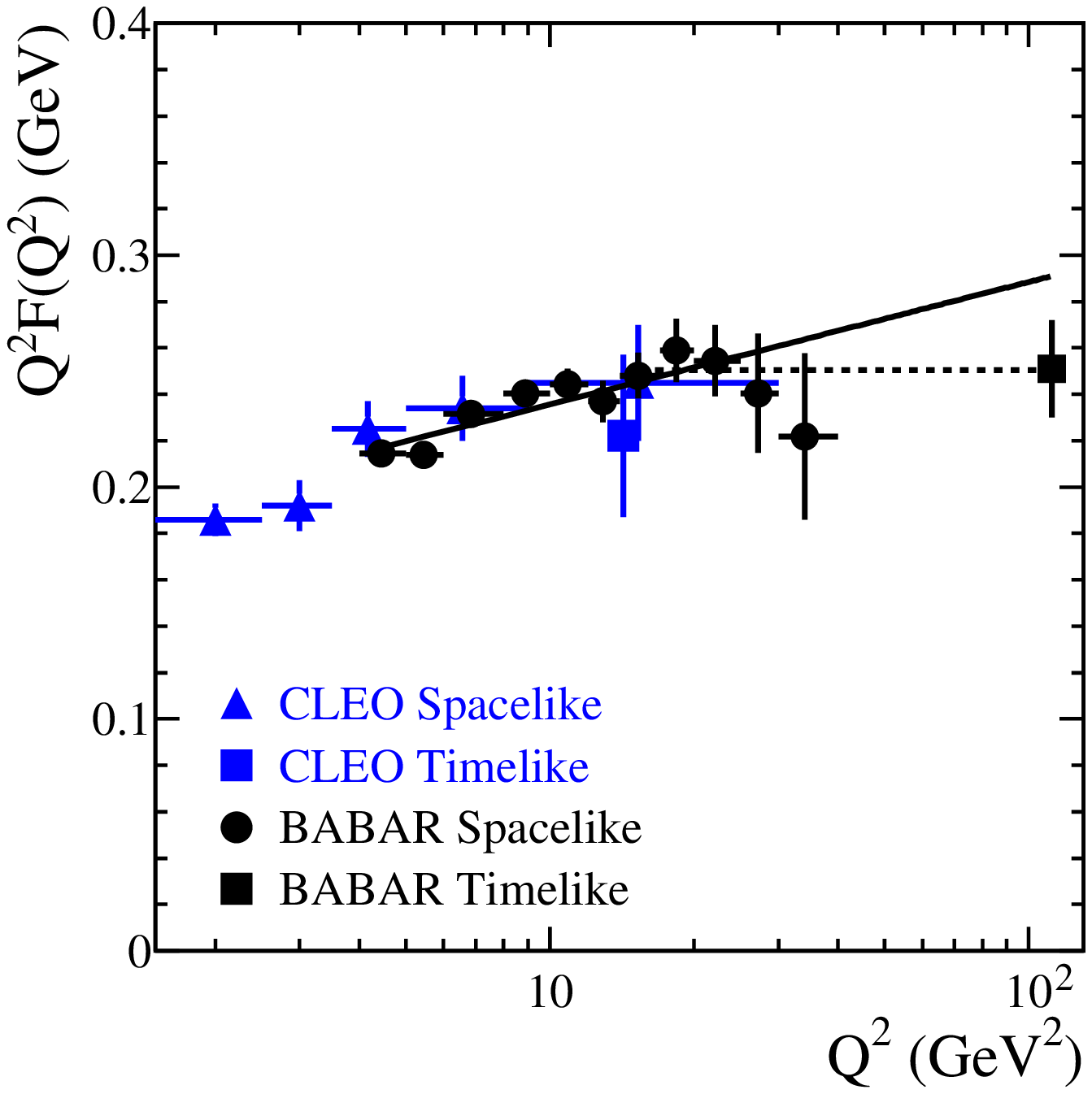}
\caption{Left:  transition form factor for the \etac\ meson as a
  function of $Q^2$, normalized to its value at $Q^2 \!=\! 0$.
  The dashed line represents the prediction of leading order QCD, 
  and the blue line the result of the fit to the monopole function
  described in the text.
  Timelike (squares) and spacelike (circles) TFFs for the \etaa\
  (middle) and \etap\ (right) mesons scaled by $Q^2$, 
  along with previous results from CLEO (blue).
  The dotted and solid lines are constant and logarithmic fits,
  respectively, to the BaBar data for $Q^2$ above 12 and 3~GeV$^2$.
} 
\label{tffeta}
\end{figure*}

For the \etac\ we also measure the cross section for untagged events, 
in which neither the $e^+$ nor the $e^-$ is detected,
and which corresponds to $F(0)$.
We use this for normalization,
and the resulting $F_{\eta_c}(Q^2)/F_{\eta_c}(0)$ is shown in
Fig.~\ref{tffeta}(left) as a function of $Q^2$.
It shows the expected falling behavior with $Q^2$, 
and the data points lie systematically below the leading order QCD
prediction (dashed red line),
indicating the need for a higher-order calculation.
A fit to a monopole function, $F(Q^2) \!=\! F(0) / (1+Q^2/\Lambda)$,
gives a good $\chi^2$ (blue line) and yields a parameter value of
$\Lambda \!=\! 8.5\pm0.6\pm0.7$~GeV$^2/c^4$,
which is consistent with a prediction based on vector dominance of
$\Lambda \!=\! m^2_{J/\psi} \!=\! 9.6$~GeV$^2/c^4$.

For the other mesons we convert our measured cross sections into TFFs
and scale by $Q^2$, since the quantity $Q^2F_X(Q^2)$ is expected to
approach an asymptotic value of $\sqrt{2}f_X$, where $f_X$ is the
$X$-meson decay constant.
There are overall systematic uncertainties of 2.3\%, 2.6\% and 3.3\% on
the \pz, \etaa\ and \etap\ data, respectively.
The resulting $Q^2F_{\eta}(Q^2)$ and $Q^2F_{\eta^\prime}(Q^2)$ are
shown in the middle and right plots of Fig.~\ref{tffeta}, respectively,
along with previous results from CLEO.
The squares indicate timelike measurements and the circles spacelike
TFFs.
The CLEO and BaBar data are consistent, as are the timelike and
spacelike results.

Both plots show the expected logarithmic rise with $Q^2$ at low $Q^2$;
at higher $Q^2$, the data are consistent with both a continued rise
and an approach to a constant value above about 10~GeV$^2$.
More precise and/or higher $Q^2$ data are needed to determine the
asymptotic behavior.
Unfortunately, the values of $f_\eta$ and $f_{\eta^\prime}$ are not
well known, as they depend strongly on the degree of mixing between
the two mesons.
The predicted asymptotic values of $Q^2F(Q^2)$ span wide ranges that
include both the highest-$Q^2$ data points and the values from the two
fits at that $Q^2$.

\begin{figure*}[ht]
\centering
\includegraphics[width=87mm]{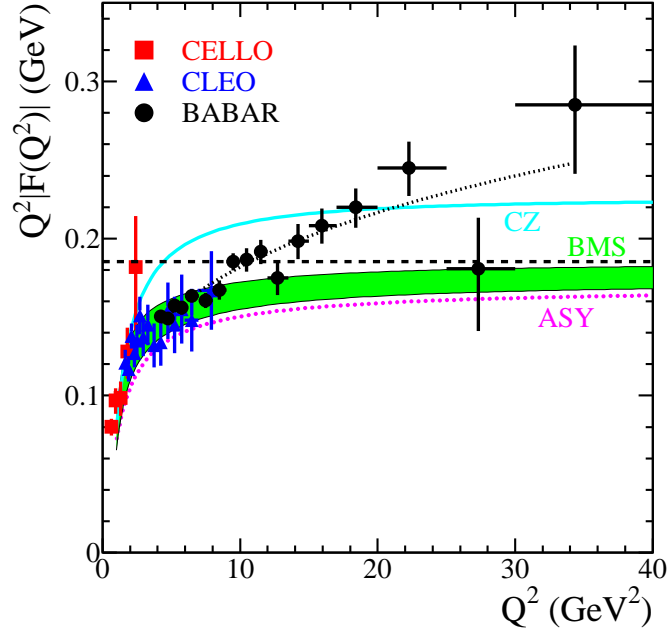}
\caption{
TFFs for the \pz\ meson scaled by $Q^2$, 
  along with previous results from CLEO (blue) and CELLO (red).
The dashed line indicates the expected asymptotic value of 0.185~GeV,
  and the dotted black line the result of a simple fit to a power of
  $Q$ (see text).
The green band and the cyan and magenta lines represent the
  theoretical predictions discussed in the text.}
\label{tffpiz}
\end{figure*}

The measured $Q^2F_\pi(Q^2)$ is shown in Fig.~\ref{tffpiz}.
It also shows the expected rise at low $Q^2$ and is consistent
with both a continued rise and a levelling off at higher $Q^2$.
However, in this case there is a firm expected asymptotic value of
$\sqrt{2}f_\pi \!=\! 0.185$~GeV,
shown by the dashed line,
and the data are above this value for $Q^2$ greater than about 10~GeV$^2$.
This indicates that the asymptotic value will be approached from above
at much higher $Q^2$ than is covered by the current measurements, if
it is approached at all,
and puts strong constraints on models for the distribution
amplitudes.

At the time of this measurement, there were few theoretical
predictions for the distribution amplitudes,
and we tested them using the formalism of Bakulev, Mikhailov and
Stefanis (BMS)~\cite{bms}.
The asymptotic form (ASY, see e.g.~\cite{bl}) gives the magenta curve on
Fig.~\ref{tffpiz};
it lies systematically below the data and shows a slow approach to the
asymptotic value from below.
The amplitude of BMS~\cite{bmsa} yields the green band on Fig.~\ref{tffpiz};
it is similar in shape to the ASY prediction but higher everywhere and
consistent with the data for $Q^2$ below 8~GeV$^2$;
however, it is inconsistent with the high-$Q^2$ data.
The amplitude of Chernyak and Zhitnitsky CZ~\cite{cz} yields the cyan
line on Fig.~\ref{tffpiz};
it is also similar in shape, but goes above the asymptotic line;
it is consistent with the data for low and high $Q^2$, but
inconsistent in shape and in the 4--14~GeV$^2$ region.
We also fitted the function 
$Q^2F(Q^2) \!=\! A(Q^2/10\,\, {\rm GeV}^2)^\beta$ to our data,
obtaining parameter values of $A \!=\! 0.182\pm 0.002$~GeV and 
$\beta \!=\! 0.25\pm 0.02$.
The latter differs significantly from the value of 0.5 predicted by
leading order QCD.

The \etaa\ and \etap\ states can be described as mixtures of 
strange ($|s\rangle \!=\! |s\bar{s}\rangle$) and nonstrange 
($|n\rangle \!=\! (|u\bar{u}\rangle + |d\overline{d}\rangle )/\sqrt{2}$)
states.
One might expect the nonstrange state to have the same decay constant
as the \pz, $f_n \!=\! f_\pi$ and an asymptotic TFF value of
$5\sqrt{2}f_n/3$, with the factor of 5/3 due to quark charges.
Similarly, for the strange component one might expect
$f_s \!=\! \sqrt{2f_K^2-f_\pi^2} \!=\! 1.36f_\pi$
and an asymptotic TFF value of $2f_s/3$.

\begin{figure*}[ht]
\centering
\includegraphics[width=84mm]{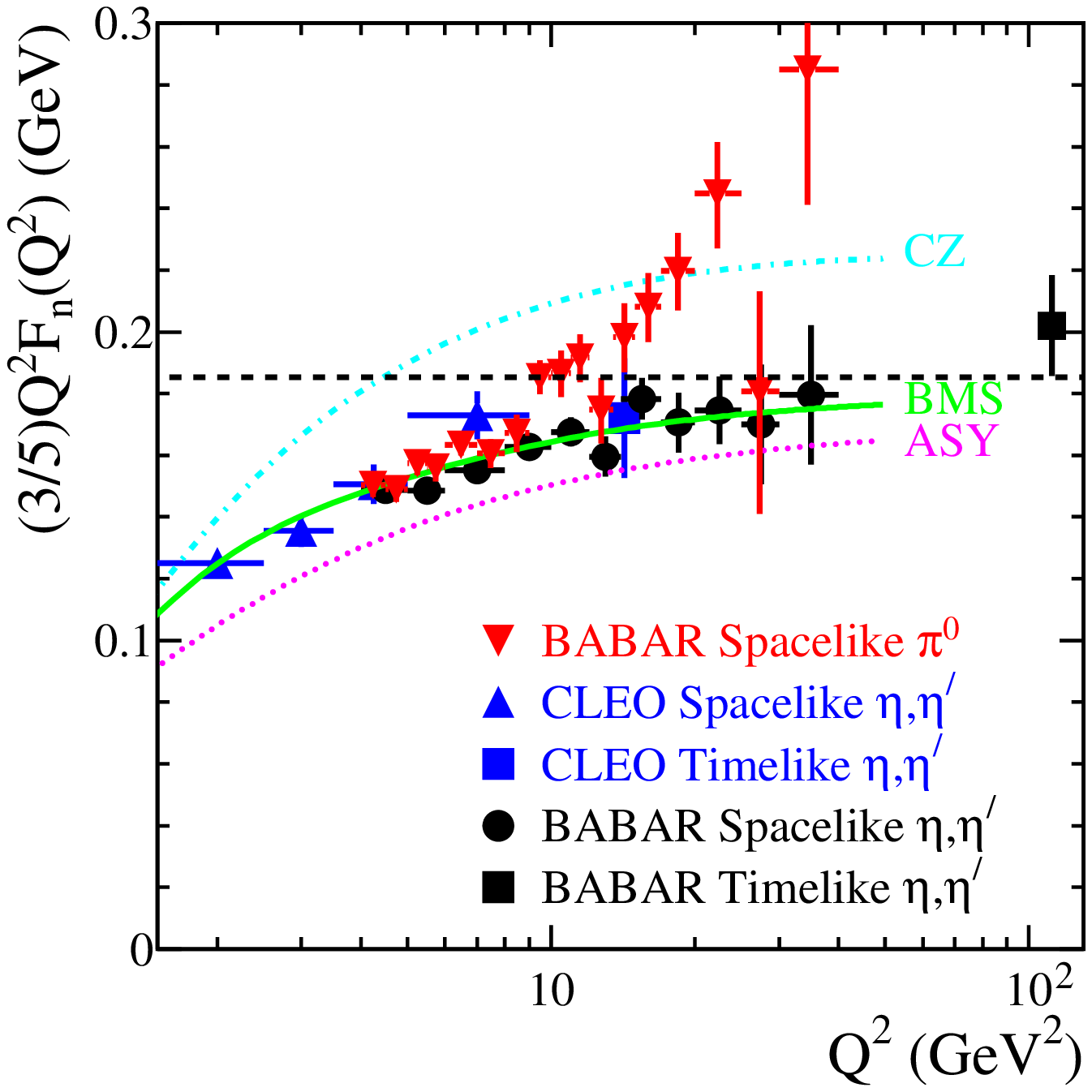}
\includegraphics[width=84mm]{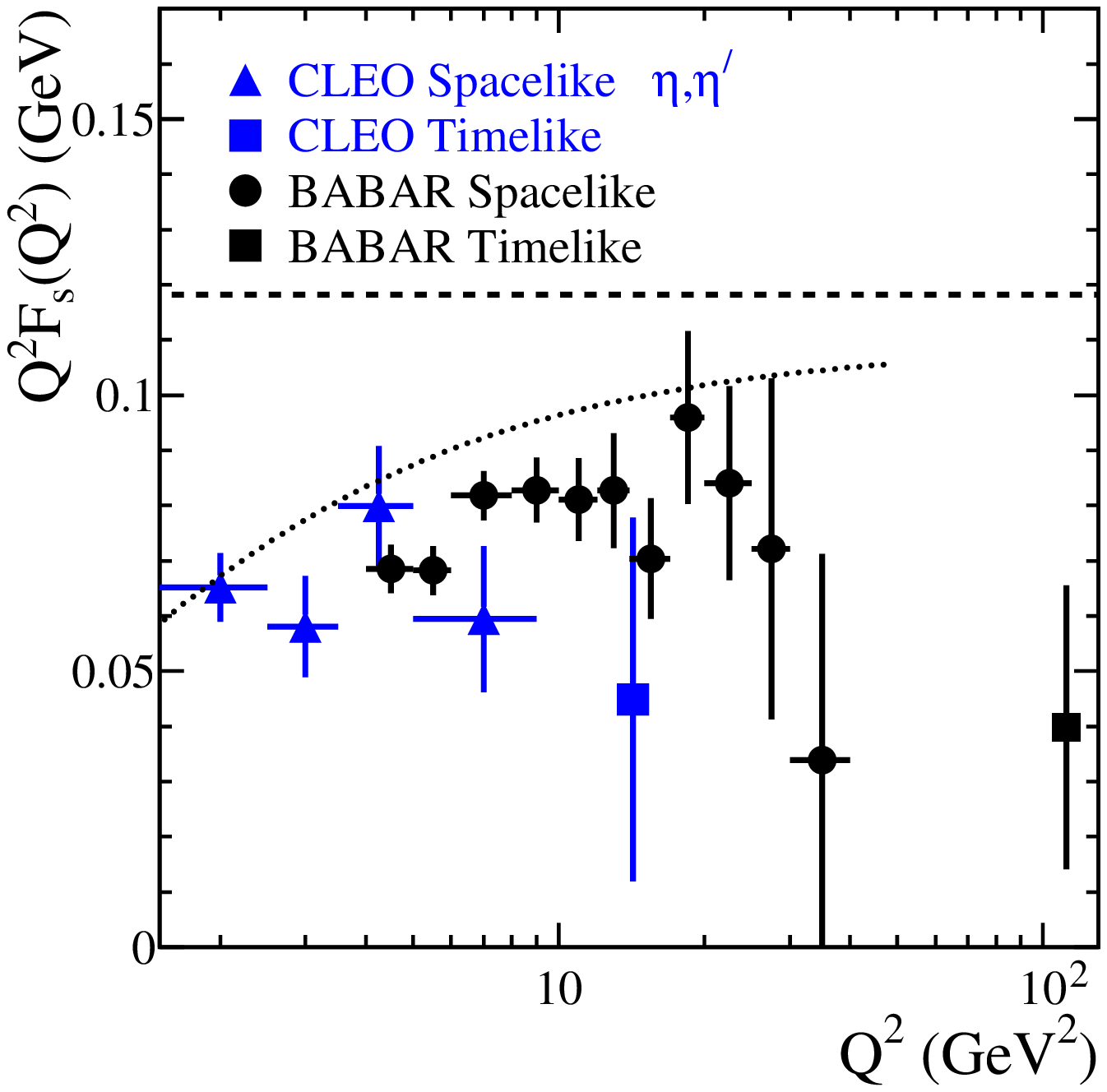}
\caption{
The nonstrange (left) and strange (right) TFFs extracted from the
\etaa\ and \etap\ data using a mixing angle of 41$^\circ$.
The nonstrange data have been scaled by three-fifths for comparison
with the \pz\ data and the theoretical predictions.}
\label{tff-pieta}
\end{figure*}

The mixing angle is not well known,
but the properties of the nonstrange component are not very sensitive
to the choice of angle.
Using a mixing angle of 41$^\circ$, 
we obtain $F_n$ and $F_s$ values from our data and the CLEO data.
The quantity $(3/5)Q^2F_n(Q^2)$ is shown in Fig.~\ref{tff-pieta}(left),
along with the BaBar $Q^2F_\pi(Q^2)$ data.
It shows the expected general behavior, 
is consistent with approaching the expected asymptotic value from
below, 
and is described very well by the BSM prediction.
The $Q^2$ dependence is also described by the ASY and CZ predictions.
However, the scaled values are inconsistent with $F_\pi$ for $Q^2$
above 10~GeV$^2$.
The strange component is shown in the right-hand plot of Fig.~\ref{tff-pieta}.
It shows similar general behavior, and the data points lie below a
simple expectation of $(\sqrt{2}/3)(f_s/f_\pi)$ times the ASY prediction
from Fig.~\ref{tffpiz}.
However,
the overall scale is sensitive to the mixing angle, 
as well as any other state mixing with the \etap,
so no conclusions can be drawn.

\section{Summary}

A very wide range of physics has been made possible by the high
luminosity of the $B$ factories, 
combined with state of the art detectors.
In the area of hadronic spectroscopy, a large number of new states has
been discovered or observed,
including expected bottomonium and charmonium states,
unexpected charmonium-like states,
charmed mesons and baryons,
the $Y(2175)$,
and new $\rho$ states.
At Babar, we have studied the production of a number of hadronic
final states via initial state radiation and two-photon interactions,
which has provided a large body of information on lighter states,
such as \etaa, $\rho$, $\omega$ and $\phi$,
as well as the production characteristics of other light mesons
and decay branching fractions of the $J/\psi$ and $\psi(2S)$ mesons.

Here we report improved measurements of the \eekkppc and \eekkppn\
cross sections, 
including a breakdown into resonant components.
In particular, we obtain much improved values for the mass, width and
production properties of the $Y(2175)$ meson.
However, its quark content and quantum numbers remain unknown.
We also report a very precise measurement of the \eepipi\ cross section
at low energies that extends the range of such measurements greatly and
and will provide a good understanding of excited $\rho$ states.
It also improves our knowledge of the total hadronic cross section,
which is vital for the interpretation of measurements of the muon
anomalous magnetic moment,

We also summarize here our measurements of transition form factors for
the pseudoscalar mesons \pz, \etaa, \etap\ and \etac.
These are consistent with previous measurements at low $Q^2$ and
extend the $Q^2$ range considerably.
The measured TFFs for the three \etaa\ states are consistent with
expectations, although higher order QCD calculations are now needed.
However, the TFF measured for the \pz\ is quite different from both
the \etaa\ mesons and the theoretical expectations at high $Q^2$.
We look forward to renewed theoretical activity on this front.

\bigskip 

\end{document}